# Adsorption and ultrafast diffusion of lithium in bilayer graphene: ab initio and kinetic Monte Carlo simulation study

Kehua Zhong,[1,2*] Ruina Hu,[1] Guigui Xu,[1,3] Yanmin Yang,[1,2] Jian-Min Zhang,[1,2]

and Zhigao Huang[1,2**]

[1]Fujian Provincial Key Laboratory of Quantum Manipulation and New Energy Materials, College of Physics and Energy, Fujian Normal University, Fuzhou 350117, China

[2]Fujian Provincial Collaborative Innovation Center for Optoelectronic Semiconductors and Efficient Devices, Xiamen, 361005, China

[3]Concord University College, Fujian Normal University, Fuzhou 350117, China

Correspondence: khzhong@fjnu.edu.cn (K.Z.); zghuang@fjnu.edu.cn (Z.H.); Telephone/Fax: 86-591-22867577 (Z.H.).

## ABSTRACT

In this work, we adopt first-principle calculations based on density functional theory and Kinetic Monte Carlo simulations to investigate the adsorption and diffusion of lithium in bilayer graphene (BLG) as anodes in lithium-ion batteries. Based on energy barriers directly obtained from first-principle calculations for single-Li and two-Li intercalated BLG, a new equation was deduced for predicting energy barriers considering Li′s interactions for multi-Li intercalated BLG. Our calculated results indicate that Li energetically prefers to intercalate within rather than adsorb outside the bilayer graphene. Additionally, lithium exists in cationic state in the bilayer graphene. More excitingly, ultrafast Li diffusion coefficient (~$10^{-5}$ cm$^2$ s$^{-1}$) within AB-stacked BLG near room temperature was obtained, which reproduces the ultrafast Li diffusion coefficient measured in recent experiment. However, ultrafast Li diffusion was not found within AA-stacked BLG near room temperature. The analyses of potential distribution indicate that the stacking structure of BLG greatly affects its height of potential well within BLG, which directly leads to the large difference in Li diffusion. Furthermore, it is found that both the interaction among Li ions and the stacking structure cause Li diffusion within AB-stacked BLG to exhibit directional preference. Finally, the temperature dependence of Li diffusion is described by the Arrhenius law. These findings suggest that the stacking structure of BLG has an important influence on Li diffusion within BLG, and changing the stacking structure of BLG is one possible way to greatly improve Li diffusion rate within BLG. At last, it is suggested that AB-stacked BLG can be an excellent candidate for anode material in Lithium-ion batteries.

# Ⅰ. INTRODUCTION

The performance of Lithium-ion batteries (LIBs) relies predominantly on the material properties of the electrodes. Advanced electrodes not only require high storage capacity but also require ultrafast charge/discharge rates, which are characterized by Lithium-ion diffusion coefficient. Carbon-based materials are at present the most commonly used negative electrode in LIBs. There are many studies on the structure of carbon atoms layer [1-3], and lithium adsorption in carbon-based materials [4-8]. Experimental results have demonstrated that graphene nanosheets have a good cyclic performance, and possess capacity up to 460 mA h g$^{-1}$ after 100 cycles [9]. Due to its high lithium storage capacity, high conductivity, and good mechanical flexibility, graphene has been regarded as a suitable candidate for electrode in LIBs [10,11]. However, the reported experimental Li diffusion coefficients span a very wide range, for example, from $10^{-16}$ cm$^2$ s$^{-1}$ to $10^{-6}$ cm$^2$ s$^{-1}$ for a variety of composite-graphite electrode architectures [12-20]. What affects Li diffusion? How to accelerate Li diffusion? These issues are not currently addressed. Therefore, in order to optimize the performance of negative electrode for LIBs, understanding the changes of electronic properties induced by Li intercalation and the diffusion kinetics properties of Li within these intercalated structures is imperative.

Reducing dimensions of material to nanoscale can effectively improve energy and power capabilities [21-23]. Now advanced experimental techniques make it possible to biuld the structure composed of a few atomic layers as the anode materials for LIBs. Bilayer graphene (BLG), which is a gapless semi-conductor similar to monolayer graphene, has regarded as a good candidate for negative electrode of LIBs [24-26]. Recent experiments have reported intercalation of Li in bilayer graphenes [26,27]. M. Kühne et al. have successfully monitored the in-plane Li diffusion kinetics within the BLG and measured ultrafast Li diffusion coefficient of up to $7 \times 10^{-5}$ cm$^2$ s$^{-1}$ at room temperature [27]. Theoretically, there have been some previous studies on Li diffusion in graphene or graphite [28-37]. Tachikawa calculated a Li diffusion coefficient (~$10^{-10}$ cm$^2$ s$^{-1}$) on a fluorine terminated graphene surface at room temperature [28]. K. Persson et al. [12] found a fast intralayer Li diffusion coefficient (~$10^{-7}$ cm$^2$ s$^{-1}$) in bulk graphite at room temperature. However, these theoretical studies of Li diffusion behavior do not consider the interaction between Li ions. In fact, the interaction between Li ions is always present, which must affect Li diffusion processes. Furthermore, ultrafast Li diffusion (~$10^{-5}$ cm$^2$ s$^{-1}$) in bilayer graphene at room temperature has been measured in recent experiments [27]. However, the result on this ultrafast diffusion has not been explained well. These inspired us to study the adsorption and diffusion kinetics of lithium in bilayer graphene.

In present work, we study Li diffusion coefficient in Li-adsorbed BLG systems by combining first-principles density functional theoretical (DFT) calculations with Kinetic Monte Carlo (KMC) simulations, which is designed to effectively simulate the time evolution of a system. Li migration energy barriers obtained from DFT are used as input parameters in KMC simulations to assess Li diffusion coefficient. However, when considering the interaction between Li ions, it is hard to get the energy barriers momentarily from DFT because they are sensitive to ion arrangement, which might change at any time in Li migration process. Therefore, in order to consider the interactions between Li ions in multi-Li adsorbed systems, we deduce a new equation which is based on the energy barriers obtained from DFT for single-Li and two-Li adsorbed systems to evaluate the migration energy barriers. Using the evaluated energy barriers into KMC simulations, Li diffusion coefficients under specific temperature and concentration are studied. The interations between Li and the BLG, the directional preference of Li diffusion, and the temperature dependence of Li diffusion are discussed.

More importantly, the experimental ultrafast Li diffusion coefficient within the BLG near room temperature is reproduced, and the cause for Li ultrafast diffusion is analyzed in detail.

## Ⅱ. METHODS

All first-principle calculations were implemented by using Vienna *ab initio* simulation package (VASP) with projector augmented wave method [39-45]. The exchange-correlation energy was calculated based on Perdew-Burke-Ernzerhof (PBE) [46] formulation of generalized gradient approximation (GGA) [39]. The optimized PBE van der Waals (optPBE-vdW) [47, 48] was used to consider the van der Waals forces between interlayer binding for bilayer graphene (BLG). The cut-off energy was chosen as 400 eV. The Monkhorst-Pack method is used to sample the $k$ points in the Brillouin zone. Our calculated lattice constant for graphene was 2.47 Å which is near the experimental value of 2.46 Å. The supercell consisting of a bilayer graphene and a vacuum separation was used to perform the convergence test of the layer size of graphene along *xy*-plane. Here, we notate the *x* and *y* axes are parallel and the *z* axis perpendicular to the graphene plane. To avert the artificial coupling role between graphene layers in two adjacent periodic images along *z*-axis, the vacuum separation of 35 Å was employed. After the test, the supercell consists of 6×6 graphene primitive cell was used to properly study lithium adsorption and diffusion in bilayer graphene. This is because in the 1~3 Li-adsorbed BLG systems studied in our work, the supercell consisting of 6×6 graphene primitive cell is large enough to make the Li atoms in the neighbouring supercells far apart, so that the overlap of their electronic states is very small and the Li-Li interaction among neighbouring supercells in the xy-plane can be neglected. The 3×3×1 Monkhorst-Pack $k$-meshes was adopted. The cut-off energy and the Monkhorst-Pack $k$-meshes were tested to ensure the accuracy of the results and they were consistent with previous studies [49]. The geometry optimizations were implemented until the force acting on each atom less than 0.02eV/Å. To study the possible Li diffusion pathways within BLG and their corresponding energy barriers, climbing image-nudged elastic band calculation (CI-NEB) [50] was performed. Based on the previous researches[50-52] and also according to the length of the pathways, three or five images were employed between two end points due to their simple pathways. Each image was relaxed until the forces on atom were less than 0.02eV/Å. In all calculations, spin polarization was included. Then these diffusion energy barriers were applied into KMC simulations to estimate the macroscopic Li diffusion coefficient in BLG. The detailed KMC algorithm follows the conventional procedure [53, 54]. In KMC simulation, the transition state theory (TST) is applied. According to TST, Li migrates from an initial site to its adjacent vacancy by experiencing a transition state, which has an energy barrier $E_m$ separating the two corresponding states before and after migration. Li migration follows the general KMC procedures: (1) Determine all possible migration sites. (2) Identify a series of the transition rates ($p_i$) for all possible migration states according to the transition rate defined as [55,56], $p_i = v^* \exp(-E_m/k_B T)$, where $v^*$ is the jump attempt frequency, which is taken to be $1 \times 10^{13}$ s$^{-1}$ [12, 33, 57]. $E_m$ is the migration energy barrier which can be obtained from the first-principle calculations. $k_B$, and $T$ denote the Boltzmann constant and temperature ($K$), respectively. (3) Calculate a accumulation $P_j = \sum_{i=1}^{j} p_i$, with $j=1,...,n$. Here, $n$ is the total number of the possible jumping sites. (4) Generate a uniform random number $u$ in a range of [0-1]. Then the $j$th jumping direction is selected according to $P_{j-1} < u P_n < P_j$. If the migration event occurs, then Li's distribution will be updated. The simulation moves into the next step. The simulation time $t$ is incremented by $\Delta t$ which is given by $\Delta t = -P_n^{-1} \ln u$. The diffusion coefficient can be decided by $D = \lim_{t \to \infty} \frac{<r(t)^2>}{4t}$ [55,56]. Where $<r(t)^2>$ is the mean squared displacement of Li. Each KMC simulation step contains $N_{Li}$ steps ($N_{Li}$ is the total number of Li atoms).

  Bilayer graphene (BLG) generally has AA and AB stacking structure, as shown in Fig.1. The single graphene size with 14×14 C atoms for BLG was used in the KMC simulations, using periodic boundary conditions for the simulation cell. The simulation temperature range is set to 263-333 K. For all simulations, 5000 MC steps are used. The first 3000 MC steps are disregarded to allow the system to relax to the equilibrium state. The diffusion coefficient is the average value of the last 2000 MC steps.

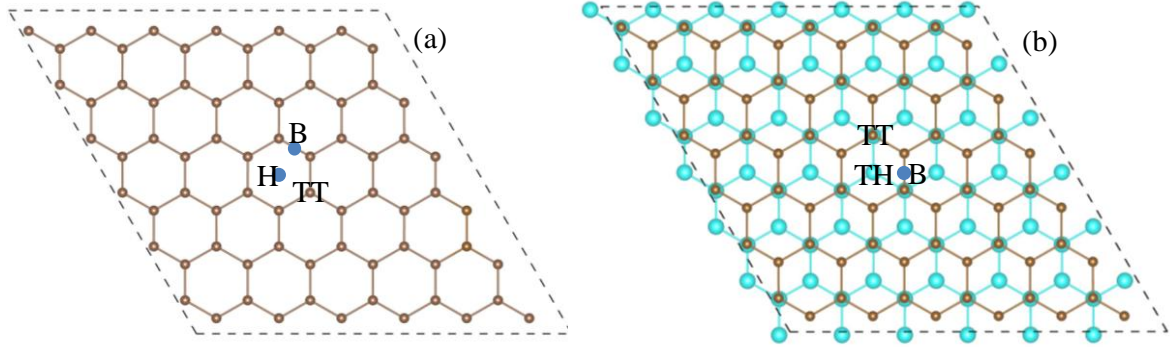

Fig.1 Top view of bilayer graphene with AA (a) and AB (b) stacking structure.

## III. RESULTS

### A. Adsorption of Li in bilayer graphene

Our calculated total energies for BLG with AA (-9.206 eV/atom) and AB (-9.210 eV/atom) stacking structure indicate that the BLG is more stable in AB stacking structure, which agrees with Refs. [58,59]. The average interlayer distances (the distance between average coordinates of carbon atoms in the upper layer and the lower layer) are 3.589 Å and 3.413 Å for AA-stacked and AB-stacked BLG, respectively. They are very close to the experimental values (3.7 Å for AA stacking and 3.35 Å for AB stacking [25]). As shown in Fig.1, there are three high symmetry sites considered for lithium adsorption within AA-stacked BLG: above two C atoms of both two graphene layers, called as top-top (*TT*) site; above the center of C hexagon or the C-C bond of two layers, called as hollow (*H*) or bridge (*B*) sites. Similarly, there are also three high symmetry sites within AB-stacked BLG: above two C atoms of both two graphene layers, called as top-top (*TT*) site; above the center of C hexagon in upper (or lower) layer, that is, the top of the C atom in lower (or upper) layer, called as top-hollow (*TH*) site; above the center of a C-C bond of upper (or lower) layer, called as bridge (*B*) site. Moreover, when lithium adsorbs outside AB-stacked or AA-stacked BLG, lithium equivalently adsorbs on a graphene monolayer in fact. Therefore, three high symmetry sites have been considered: above C atom, the center of C hexagon and C-C bond of graphene monolayer, called as top (*T*), hollow (*H*) and bridge (*B*) sites, respectively. The stability of a Li-adsorbed BLG system is generally described by the adsorption energy $E_{ad}$: [31]

$$E_{ad} = (E_{Li+BLG} - E_{Li} - E_{BLG}) . \qquad (1)$$

Where $E_{Li+BLG}$ represents the total energy for Li-adsorbed BLG system with a Li atom adsorbed in BLG. $E_{Li}$ and $E_{BLG}$ represent the energies of an isolated Li atom and the isolated BLG, respectively. The calculated adsorption energies of various Li-adsorbed BLG systems show that the favorable site locates at the hollow (*H*) one regardless of Li adsorbing within or outside BLG for AA stacking, and the top-hollow (*TH*) or hollow (*H*) one for AB stacking, as presented in Table 1. Moreover, all adsorption energies for those systems with Li intercalating within BLG are more negative than that for Li adsorbing outside BLG. Therefore, Li energetically prefers to intercalate within BLG. This finding agrees with the experimental result [27] and the first-principles calculations result [25] that Li resides only in between pristine graphene sheets. Furthermore, our calculated results also indicate that Li is more inclined to intercalate within AA-stacked BLG than AB-stacked BLG, which is consistent with previous studies [58].

Table 1. Adsorption energies $E_{ad}$ for the systems with one Li atom adsorbing within AA-stacked and AB-stacked BLG ($E_{ad,in}$) or outside AA-stacked and AB-stacked BLG ($E_{ad,out}$) at various adsorbtion sites.

| | Adsorption Site | Adsorption Energy $E_{ad}$ (meV/Å$^2$) | | Energy difference $\Delta E_{ad} = E_{ad,out} - E_{ad,in}$ (meV/Å$^2$) |
|---|---|---|---|---|
| | | $E_{ad,in}$ for one Li within BLG | $E_{ad,out}$ for one Li outside BLG | |
| AA | H | -11.34 | -6.45 | 4.89 |
|  | B | -9.55 | -5.08 | 4.47 |
|  | TT | -8.91 | -5.01 | 3.90 |
| AB | TH (within) or H (outside) | -10.66 | -6.47 | 4.19 |
|  | B | -9.94 | -5.17 | 4.77 |
|  | TT | -8.37 | -5.09 | 3.28 |

Due to a large difference in electronegativity between Li and C atoms, there is a charge transfer from Li to C atoms. The interaction between Li and C atoms can be visually explored by investigating the valence electron density distributions and the electron density difference, which are represented in Fig.2. The electron density difference is usually defined as $\Delta\rho = \rho_{Li+BLG} - \rho_{Li} - \rho_{BLG}$. Where $\rho_{Li+BLG}$, $\rho_{Li}$ and $\rho_{BLG}$ denote the electron density distributions of the Li-adsorbed BLG system, the isolated Li atom and the isolated BLG, respectively. Our calculated results of electron distributions for Li adsorbing within (or outside) AA-stacked BLG or AB-stacked BLG are the similar. All indicate that BLG is the major charge acceptor. For Li intercalating within AB-stacked BLG, as can be seen in Fig. 2(a), the electron distributions surround Li atom are rare, indicating that charge depletion takes place in this region. It suggests that Li atom donates its $2S^1$ electron to the graphene layers and exists in cationic state, which is consistent with previous reports [58,60]. Figs.2 (b)-(d) show that the charge transfers from Li atom to its nearest neighbor C atoms of BLG. The red and yellow isosurfaces denote the charge depletion and charge accumulation, respectively. Brown and turquoise balls are the C atoms located in the upper and lower layer, respecitvely. Green ball is Li atom. For Li intercalating within AA-stacked BLG (Figs. 2(g) and (h)), the behavior of the charge transfer is the same. But for the case of Li adsorbing outside AB-stacked BLG (Figs. 2(e) and (f)), the charge transfer occurs between Li and C atoms only in Li's nearest neighbor graphene layer not in the layer far away. The case for Li adsorbing outside AA-stacked is the same, so its electron density difference isn't displayed repeatedly here. When Li intercalation content increases, it would appear Li atoms being close together. Fig. 3 shows the valence electron density distributions and the electron density difference for two Li atoms intercalating closely within AB-stacked BLG. From the figure, it is clearly seen that the charge transfer only occurs rather locally surround each Li atom, which is similar to the behavior of single Li atom. To quantify the amount of electron transfer, Bader charge analysis [61] was also carried out. Our Bader charge analysis for the system of Li adsorbing within AB-stacked BLG shows that about 0.85e per Li atom is transferred from Li to the BLG, which agrees with previous reports [58,60]. Therefore, when Li atoms intercalate within BLG, we can reasonably think that they exist as Li ions.

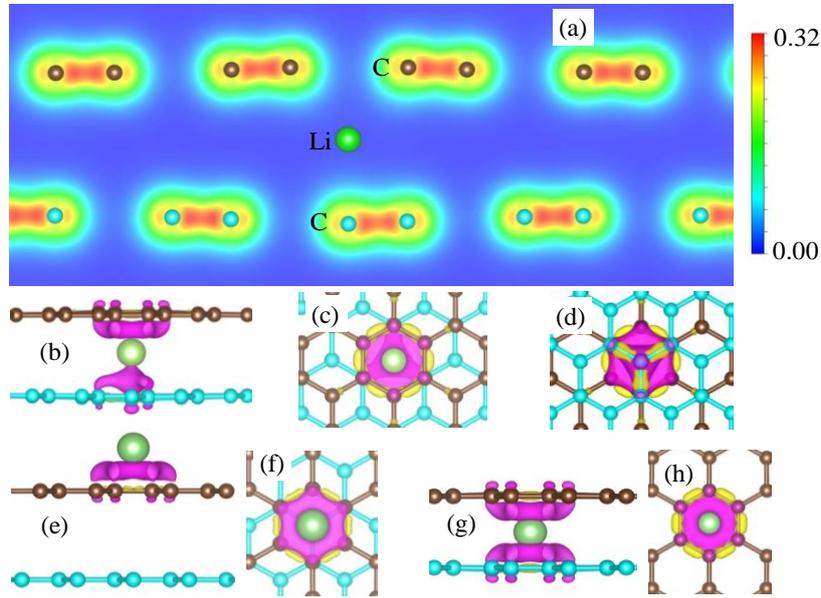

Fig. 2 (a) The valence electron density (e·bohr$^{-3}$) distributions for Li adsorbing at TH site within AB-stacked BLG. The electron density difference Δρ (b) side view, (c) top view and (d) bottom view for Li adsorbing within AB-stacked BLG, (e) side view and (f) top view for Li adsorbing at $H$ site outside AB-stacked BLG, (g) side view and (h) top view for Li adsorbing at $H$ site within AA-stacked BLG.

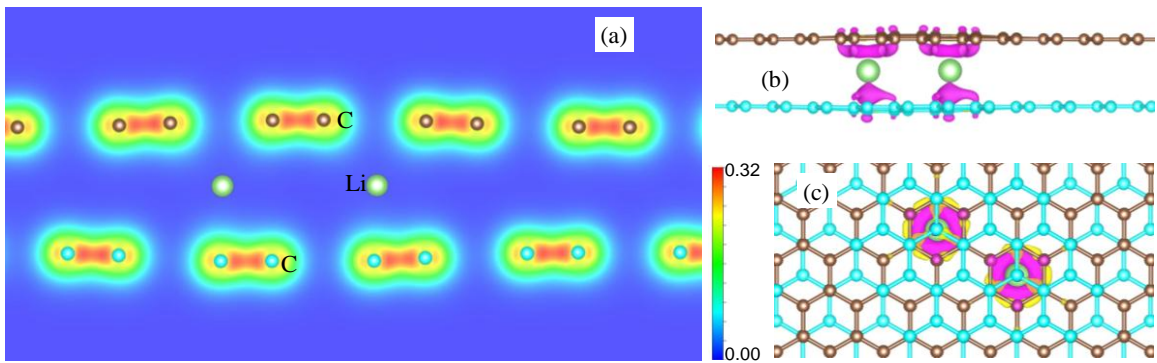

Fig.3 (a) The valence electron density (e·bohr$^{-3}$) distributions for two adjacent Li atoms adsorbing at $TH$ sites within AB-stacked BLG. The electron density difference Δρ (b) side view and (c) top view.

**B. Migration energy barrier in single-Li and two-Li adsorbed BLG systems**

Li diffusion energy barrier is an important factor of kinetics linked to the Li diffusion in intercalation electrodes for LIBs. The charge/discharge rates of metal-ion batteries predominantly depend on Li diffusion in anode materials. Therefore, it is very important to study Li diffusion in BLG intercalation electrode. The diffusion energy barriers of Li in BLG can be obtained from VASP with CI-NEB along diffusion pathways. For the system with a single Li atom intercalating within AB-stacked BLG, since the favorable site for Li locates at the *TH* site, the possible transition path that Li migrates via jumping between neighboring TH sites ($TH_1 \leftrightarrow TH_2$) was considered, as shown in Fig. 4(a). For the system with a single Li adsorbing outside AB-stacked BLG, actually the favorable site for Li locates at the H site of upper (or lower) monolayer graphene. Hence, Li migrates between the neighboring H sites ($H_1 \leftrightarrow H_2$), as shown in Fig. 4(b). The total energy profiles along corresponding diffusion pathways were also shown in Fig.4. From Fig. 4(a), a Li atom intercalated within AB-stacked BLG prefers to diffuse along $TH_1 \rightarrow TH_2$ pathway because of an obvious lower energy barrier ($\Delta E$) merely of 0.07 eV, which indicates that Li could diffuse very fast within BLG between neighboring *TH* sites. However, the energy barrier that for Li migrating between two neighboring *H* sites outside AB-stacked BLG reaches up to 0.25 eV, shown in Fig. 4(b). This clearly shows that Li diffusion within AB-stacked BLG is energetically favoured comparing with the case for Li diffusion outside BLG, which is consistent with the recent experimental finding that Li diffuses in between the graphene sheets but not on top of the bilayer graphene that is the graphene/vacuum interface [27]. For the systems with a single Li atom intercalating within and adsorbing outside AA-stacked BLG, as shown in Fig. 4(c) and (d) respectively, Li also migrates between neighboring *H* sites ($H_1$-$H_2$). The energy barrier for Li migrates between two neighboring *H* sites outside AA-stacked BLG is basically the same as that for AB-stacked BLG. Notably, however, it is obviously more difficult for Li to diffuse within AA-stacked BLG than outside AA-stacked BLG (energy barriers $\Delta E$=0.34 eV and 0.26 eV for Li migrating within and outside BLG, respectively). Furthermore, the comparison between two energy barriers of 0.07 eV and 0.34eV for Li diffusing within AB-stacked and AA-stacked BLG, interestingly, reveals that Li apparently rapidly migrates only within AB-stacked BLG but not AA-stacked BLG. Our results clearly indicate that the stacking sequence of two graphene layers can evidently impact the intercalation of Li and its diffusion in these BLG structures. Besides, it can also be found from Fig. 4(a) that along $TH_1 \rightarrow TT$ pathway Li needs to overcome a large energy barrier of 0.25 eV, while it can easily jump from $TH_1$ site to $TH_2$ site just getting over a small energy barrier of

0.07 eV. This obviously indicates that Li diffusion within AB-stacked BLG exhibits directional preference, and it may be attributed to its AB stacking structure of the BLG.

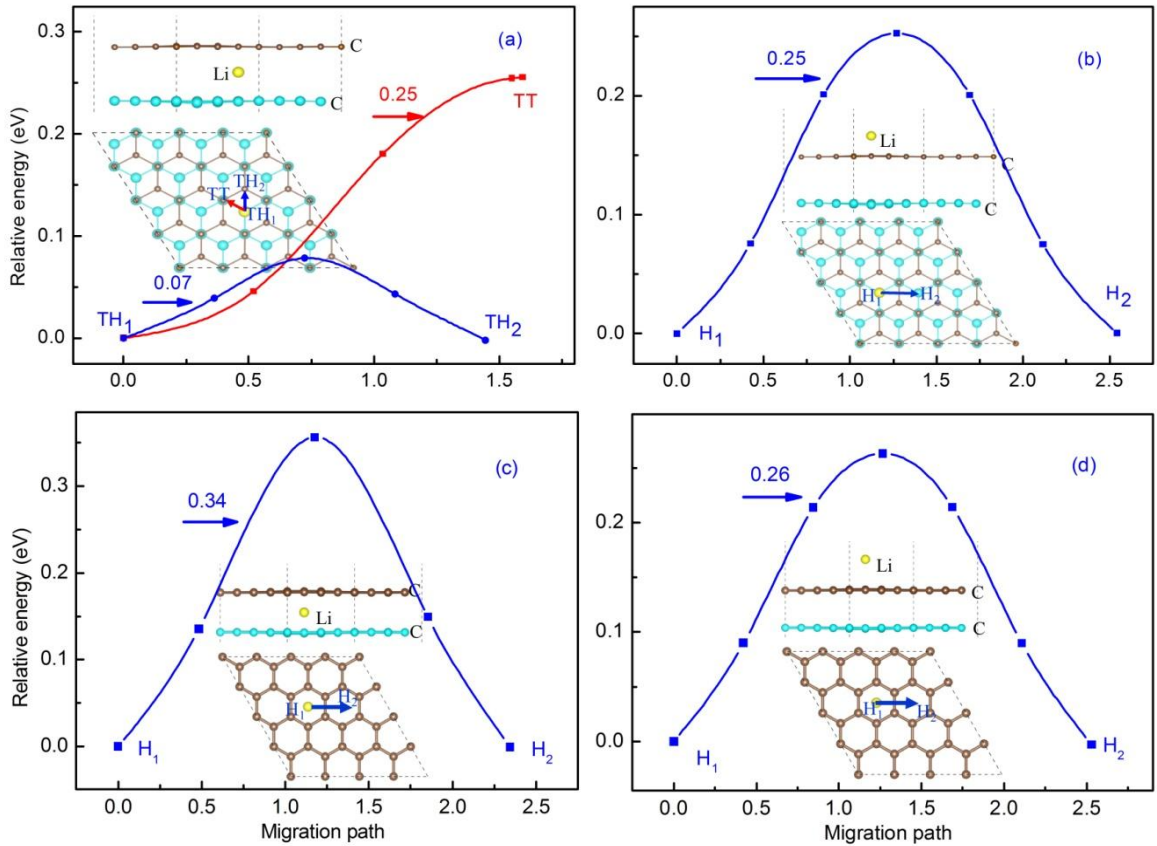

Fig. 4 The total energy profiles for a Li atom intercalated (a) within and (b) outside AB-stacked BLG, (c) within and (d) outside AA-stacked BLG.

In order to further understand the really large difference between two energy barriers for Li diffusing within AB-stacked and AA-stacked BLG and the directional preference of Li diffusion within AB-stacked BLG, the electrostatic potential distributions of the systems have been explored. For graphene monolayer, one observable feature of the electrostatic potential is its having potential well near the hollow region, as shown in Fig. 5(a). For AA-stacked or AB-stacked bilayer graphene, the electrostatic potential within them distributes periodically attributing to their own periodic lattice structure, shown in Fig. 5(b) and (d). It can be considered that the underlying graphene layer provides a periodic potential for the upper layer. The difference is that for AA-stacked BLG the potential wells generated by the upper and lower layers are the mirror images and just reside in the hollow regions of bilayer graphene, whereas for AB-stacked BLG the potential wells generated by both layers are staggered. As can be seen from Fig. 5(d), the potential wells for one layer deviate from the hollow regions of the other layer and face the C atoms of the other layer. This has resulted in that the potential wells within AA-stacked BLG are deeper than that within AB-stacked BLG, as can be observed in Fig. 5(c) and (e). This may explain why there is a large difference between two energy barriers for Li diffusion within AA-stacked and AB-stacked BLG. With intercalation of Li, the interaction between Li and C atoms greatly changes the electrostatic potential distribution near Li. Our previous finding has shown that Li diffusion within AB-stacked BLG has directional preference, so here we explain this phenomenon qualitatively. As can be seen from Fig.5 (h) and (i), the potential distribution between Li and the upper layer is approximately the same in the *TT* and *TH* directions. However, in the region between Li and the lower layer, the potential distribution (Fig.5 (h) and (j)) is obviously different in the two directions. The potential distribution along *TH* direction would help Li easier to jump from the present potential well to another nearest potential well along this direction. This corresponds to the previous result that the energy barriers of Li diffusion along $TH_1 \rightarrow TH_2$ pathway is lower than that along $TH_1 \rightarrow TT$ pathway within AB-stacked BLG.

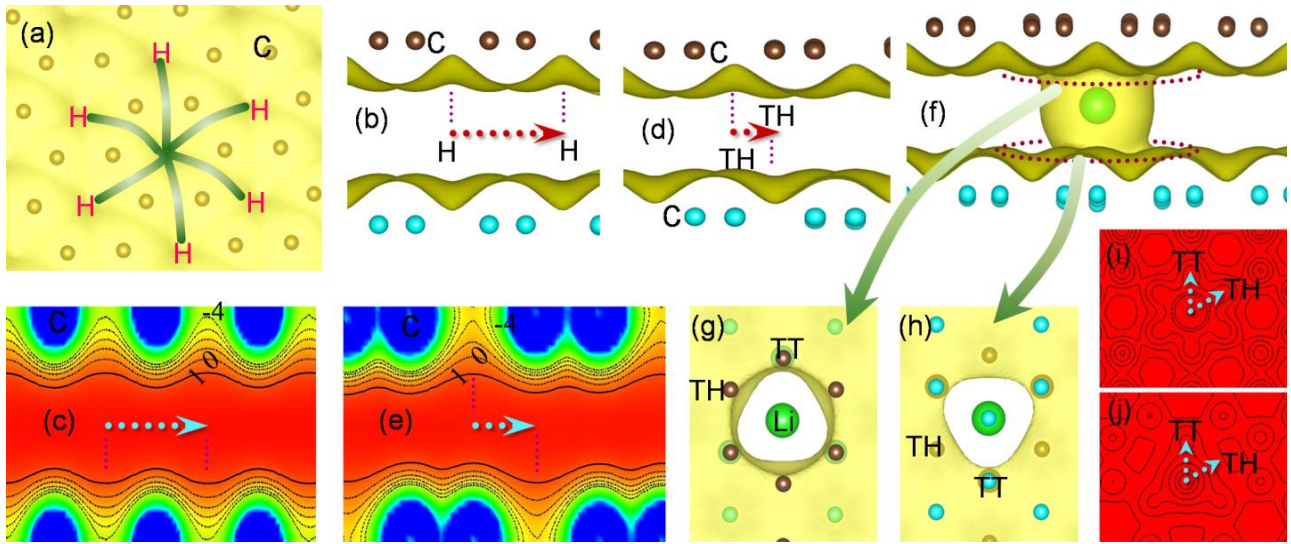

Fig. 5 The side view for the electrostatic potential (eV) isosurface for graphene monolayer (a) and within (b) AA-stacked BLG and (d) AB-stacked BLG. (c) and (e) Respective contour and profile maps of electrostatic potential along the directions of $H \rightarrow H$ and $TH \rightarrow TH$ paths for (b) and (d). The side view for the electrostatic potential isosurface of the system for Li-adsorbed within AB-stacked BLG (f), and the corresponding top view (g) and bottom view (h). The contour and profile maps between Li and upper graphene (i), between Li and lower graphene (j). The isosurface level is -0.23 eV.

According to the above findings: Li atom becomes Li ion when it intercalates into the BLG. Therefore, when two Li ions are close together, they would interact with each other. This interaction would have an important influence on the electronic structure and the related properties of the system, such as Li diffusion energy barrier in BLG. So, for rare Li intercalated into BLG, Li diffusion can be seen as the migration behavior for non-interacting particles. While Li content increases, ions being close together will lead to the inevitable interaction between Li ions. Moreover, Li energetically prefers to intercalate within BLG rather than adsorb outside BLG for both AA-stacked and AB-stacked BLG, hence only the cases for Li intercalating within BLG are considered in the following analyses of the interaction between Li ions. To study the interaction between Li ions, we calculated the total energy profiles along possible migration pathways for the two Li ions (named as two-Li) intercalated BLG systems that $Li_1$ is fixed while $Li_2$ is migrating, as shown in Fig. 6. For two-Li intercalated within AA-stacked BLG, as shown in Fig. 6(a), theoretically when $Li_1$ already exists, the nearest neighbor site for $Li_2$ should only locate at $H_2$ site. It can also be confirmed in the total energy profile shown in Fig.6 (a). From the figure, the closer neighbor H-site $H_4$ of $Li_1$ is the energetically unstable site for $Li_2$ due to its obviously higher energy than those of other H-sites ($H_1$, $H_2$, $H_3$). Thus, the $H_4$ site acts as a transition state, and if $Li_2$ locates at $H_4$ site, it could easily hop into its neighbor lower energy sites $H_2$ or $H_3$ via overcoming a lower energy barrier about 0.10 eV relative to the energy barrier of 0.40eV for its reverse migration from $H_3$ site to $H_4$ site. In Fig.6 (a), three possible migration pathways (labled as $H_1 \rightarrow H_2$, $H_3 \rightarrow H_2$, $H_3 \rightarrow H_4$) for $Li_2$ ion in its migration behavior were investigated, assuming another $Li_1$ ion has already existed on $Li_2$'s theoretically nearest-neighbor site. In general, $Li_2$ demands relatively larger energy barriers to jump between $H_2$ and $H_1$ sites or $H_3$ and $H_2$ sites. According to our calculations, $Li_2$ needs to get over energy barriers about 0.32 eV and 0.33 eV for pathways $H_1 \rightarrow H_2$ and $H_3 \rightarrow H_2$, respectively. In the reverse migration processes, $Li_2$'s migration also requires overcoming large barriers of 0.27 eV and 0.30 eV for pathways $H_2 \rightarrow H_1$ and $H_2 \rightarrow H_3$, respectively. Notwithstanding, $Li_2$'s reverse migration processes still surpass the relatively lower barriers. Moreover, the differences of Li diffusion barriers along the paths $H_2 \rightarrow H_1$ and $H_2 \rightarrow H_3$ are observed. These may be attribute to the interaction between close Li ions. In other words, the interaction between Li ions causes Li diffusion within AA-stacked BLG to have directional preference.

For two-Li intercalated within AB-stacked BLG system, as shown in Fig. 6(b), when $Li_1$ already exists, the theoretically nearest neighbor site for $Li_2$ should locate at $S_2$ (or $S_5$) TH-site. The closer neighbor TH-site $S_4$ of $Li_1$ is energetically more unstable for $Li_2$ than other TH-sites ($S_1$, $S_2$, $S_3$, $S_5$, $S_7$, $S_9$). Thus, the $S_4$ site acts as a transition state. If $Li_2$ locates at $S_4$ site, it could easily jump into its adjacent lower energy sites $S_3$ or $S_5$ via overcoming a very low energy barrier about 0.03 eV. Furthermore, from the figure, the much higher energies of those TT-sites once again prove their instability, and also confirmed that Li is difficult to migrate along their corresponding pathways. That is, Li migrating from TH-site to TH-site is significantly easier than migrating from TH-site to TT-site. On the other hand, the barriers (0.07eV, 0.07eV, 0.08eV and 0.06eV) along pathways $S_2 \rightarrow S_1$, $S_7 \rightarrow S_3$, $S_9 \rightarrow S_2$ and $S_5 \rightarrow S_4$ are slightly higher than those barriers (0.05eV, 0.06eV, 0.07eV and 0.03eV) along their reverse migration pathways $S_1 \rightarrow S_2$, $S_3 \rightarrow S_7$, $S_2 \rightarrow S_9$ and $S_4 \rightarrow S_5$. Just as the case for AA-stacked

BLG, the Li-Li repulsion leads Li diffusion to have directional preference within AB-stacked BLG, although the impact is smaller than that induced by the AB stacking structure. Therefore, AB stacking structure and Li-Li repulsion are the main contributing factors that influence Li diffusion within AB-stacked BLG to exhibit directional preference.

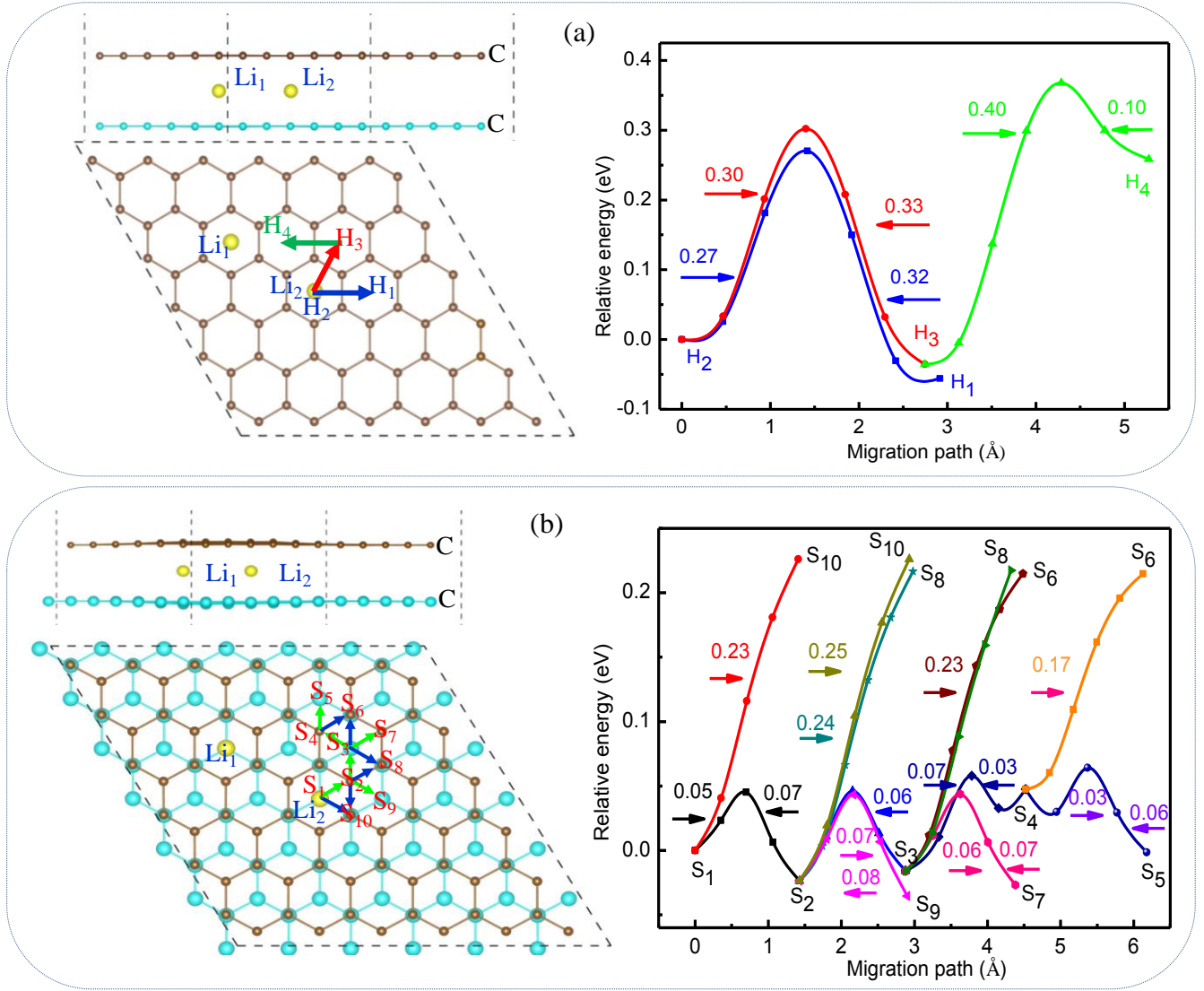

Fig. 6 The total energy profiles for two-Li intercalated within (a) AA-stacked BLG and (b) AB-stacked BLG.

## C. Li diffusion in multi-Li intercalated BLG

Macroscopic Li diffusion coefficient $D_{Li}$ is an important physical quantity used to characterize the speed of Li diffusion, and it can be estimated from the migration energy barriers. The previous report has shown that Li may intercalate into bilayer graphene up to a stoichiometry of $C_6LiC_6$ [26]. That is, the maximum Li content in $Li_xC_{12}$ can only reach 1. For our studied AA-stacked and AB-stacked BLG systems with very low Li content x such as x=0.03 and 0.11, since in such systems Li-Li interactions are hardly considered, we can estimate Li diffusion coefficients $D_{Li}$ for these $Li_xC_{12}$ BLG systems based on our previous migration energy barriers (as shown in Fig. 4(a), (c) and Fig. 6). Our calculated Li diffusion coefficients $D_{Li}$ for $Li_xC_{12}$ with very low Li content x=0.03 and 0.11 near room temperature (300 K) are about $0.16\times10^{-6}$ and $0.3\times10^{-7}$ $cm^2 \cdot s^{-1}$ for AA-stacked BLG systems, while they are about $0.94\times10^{-3}$ and $0.16\times10^{-3}$ $cm^2 \cdot s^{-1}$ for AB-stacked BLG systems. It suggests that sparse Li atoms could ultra-fast diffuse within AB-stacked BLG instead of AA-stacked BLG. The recent experiment has measured an ultrafast diffusion of Li within bilayer graphene at room-temperature [27], so we only consider the systems of Li-intercalated within AB-stacked BLG in the following study of the influence of Li-Li repulsion on Li's diffusion kinetics. Moreover, Sharmila et al. [58, 59] found that for AB-stacked bilayer graphene with the Li/C ratio being more than 1/12, AB-stacked structure of layers would switch into AA-stacked structure. For higher Li intercalation content (Li/C ratio being more than 1/12), the Li-adsorbed BLG became energetically unstable which was manifested by positive adsorption energies [58]. Similarly, our results for two-Li intercalated within AB-stacked BLG show that with the presence of $Li_1$ the closest stable site for $Li_2$ locates at $S_1$ TH-site. As Li intercalation content increasing, when all such closest stable sites are filled, it corresponds to that the maximum x for $Li_xC_{12}$ is 1. Therefore, in the following study on the influence of interactions among Li ions on their diffusion kinetics, only those $Li_xC_{12}$ systems with Li content x<1 for Li-intercalated within AB-stacked BLG are considered.

For the systems of rare Li intercalated into AB-stacked BLG, Li ions are far away from each other. Therefore, Li-Li interactions through Coulomb repulsion are negligible. Li diffusion can be seen as the migration behavior for non-interacting particles. Thus Li diffusion coefficients may be easily estimated based on those migration barriers directly obtained from First-principle calculations combined with CI-NEB method. However, for multi-Li intercalated AB-stacked BLG with higher Li content, Li ions being close enough would interact with each other through Coulomb repulsion, and then play an important role on Li's migration energy barriers, ultimately affect Li's diffusion process within BLG. Considering the interactions among many ions would have the problem become very complicated. Li's arrangements constantly change during diffusion process, then Li's migration barriers change subsequently. Therefore, it is difficult to describe the influence of interaction between Li ions on the energy barrier using first-principle calculations. To further study the influence of Li's content on diffusion behavior, we have adopted an equation which has been developed to investigate the influence of Li's interactions on their migration energy barriers for multi-Li adsorbed graphene in our previous work [62] to describe the interactions of Li ions intercalated within AB-stacked BLG. In the equation, Li migration barrier of a multi-Li adsorbed system was simplified into a functional relation of the Li migration barriers of one-Li and two-Li adsorbed systems. These Li migration barriers for one-Li and two-Li adsorbed systems are directly calculated based on First-principle calculations combined with CI-NEB method. The equation is briefly derived in the next section.

A schematic diagram for multi-Li intercalated AB-stacked BLG is illustrated in Fig. 7. The energies for *a* and *c* sites on migration path can be estimated as

$$E_a = K_1 \left( \frac{1}{r_{0,1}} + \frac{1}{r_{0,2}} + \frac{1}{r_{0,3}} + \cdots + \frac{1}{r_{0,i}} \right) + K_2 E_{0,\text{BLG}} + H_0, \tag{2}$$

$$E_c = K_1 \left( \frac{1}{r'_{0,1}} + \frac{1}{r'_{0,2}} + \frac{1}{r'_{0,3}} + \cdots + \frac{1}{r'_{0,i}} \right) + K_2 E'_{0,\text{BLG}} + H_0. \tag{3}$$

Where $E_a$ ($E_c$) denotes the total energy for multi-Li intercalated AB-stacked BLG with migrated Li (labeled as the *0*-th Li$_0$) locating at *a* (or *c*) site. $E_{0,\text{BLG}}$ (or $E'_{0,\text{BLG}}$) is the energy for the system which includes only the *0*-th Li locating at *a* (or *c*) site within BLG. $H_0$ denotes the total energy for multi-Li intercalated AB-stacked BLG with the *0*-th Li removed. $K_1$ and $K_2$ are proportionality coefficients. $r_{0,i}$ and $r'_{0,i}$ denote the distances between the *0*-th and *i*-th Li.

Subtracted Eq. (3) from Eq. (2), and eliminated the same $H_0$ term in the two equations, the energy difference $\Delta E_{ac}$ between *a* and *c* sites (corresponding to initial state and activated state, respectively) on the migration path $a \rightarrow c \rightarrow b$ can be estimated as

$$\Delta E_{ac} = E_a - E_c = \left[ K_1 \left( \frac{1}{r_{0,1}} + \frac{1}{r_{0,2}} + \frac{1}{r_{0,3}} + \cdots + \frac{1}{r_{0,i}} \right) + K_2 E_{0,\text{BLG}} \right]$$

$$- \left[ K_1 \left( \frac{1}{r'_{0,1}} + \frac{1}{r'_{0,2}} + \frac{1}{r'_{0,3}} + \cdots + \frac{1}{r'_{0,i}} \right) + K_2 E'_{0,\text{BLG}} \right]. \tag{4}$$

Subtract the corresponding items of the two terms of Eq. (4), the energy difference $\Delta E_{ac}$ can be rewritten as

$$\Delta E_{ac} = K_1 \left[\left(\frac{1}{r_{0,1}} - \frac{1}{r'_{0,1}}\right) + \left(\frac{1}{r_{0,2}} - \frac{1}{r'_{0,2}}\right) + \cdots + \left(\frac{1}{r_{0,i}} - \frac{1}{r'_{0,i}}\right)\right] + K_2(E_{0,BLG} - E'_{0,BLG}). \quad (5)$$

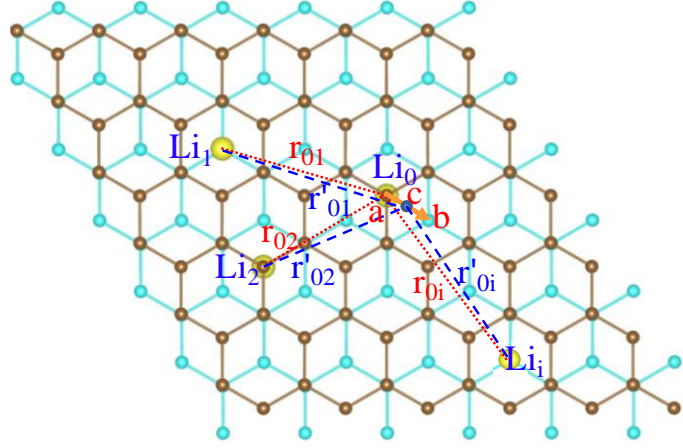

Fig. 7 A schematic diagram for multi-Li intercalated AB-stacked BLG.

For a two-Li intercalated BLG system, such as the system with only two Li atoms (*0*-th and *i*-th) intercalating within BLG, the energies for the *0*-th Li locating at *a* site $E_{0,i}$ and for the *0*-th Li locating at *c* sites $E'_{0,i}$ can be estimated as

$$E_{0,i} = K_1 \frac{1}{r_{0,i}} + K_2 E_{0,BLG} + H'_0, \quad (6)$$

$$E'_{0,i} = K_1 \frac{1}{r'_{0,i}} + K_2 E'_{0,BLG} + H'_0. \quad (7)$$

Where $H'_0$ denotes the total energy for the two-Li intercalated AB-stacked BLG with the *0*-th Li removed. The energy difference $\Delta E_{0,i}$ can be estimated as

$$\Delta E_{0,i} = E_{0,i} - E'_{0,i} = K_1 \left(\frac{1}{r_{0,i}} - \frac{1}{r'_{0,i}}\right) + K_2(E_{0,BLG} - E'_{0,BLG}). \quad (8)$$

Combined Eq. (8) with Eq. (5), the energy difference for multi-Li intercalated BLG system can be derived as

$$\Delta E_{ac} = (\Delta E_{0,1} + \Delta E_{0,2} + \cdots + \Delta E_{0,i}) - (i-1) \times \Delta E_{0,BLG}. \quad (9)$$

Here, $\Delta E_{0,BLG} = K_2(E_{0,BLG} - E'_{0,BLG})$ is the energy difference for one-Li intercalated BLG system with single Li migrating from *a* to *c* site, which is the diffusion barrier for single-Li intercalated BLG system. In Eq. (9), the first term includes the energy differences corresponding to those two-Li intercalated BLG systems. The second term is merely related to the single-Li intercalated BLG system without interaction between Li ions. Therefore, the diffusion barriers of multi-Li intercalated BLG system can be simplistically estimated from diffusion barriers of two-Li and single-Li intercalated BLG systems. To check again our method for energy barrier estimation, we used two methods to estimate the energy barriers along various possible migration pathways for three-Li intercalated BLG system. One is that estimating the energy barriers from the total energy curves obtained by first-principle calculations combined with CI-NEB method. Fig. 8(a) shows the total energy curves along migration pathways corresponding to Fig. 8(b) by CI-NEB method. The energy barriers of the migration pathways were also marked in the figure. The other method is to estimate the energy barriers directly by Eq. (9) based on the total energy curves of single-Li and two-Li intercalated BLG systems obtained from first-principles calculations. The results from two methods are presented in Fig. 9. Just as the coincidence has already been obtained in our previous study for multi-Li adsorbed graphene system [62], it can be observed once again that the results from both methods are consistent well. It can also be described by the mean deviation Δ calculated as follows:

$$\Delta = \frac{1}{N}\sum_{i=1}^{N} \frac{|D_{i,ab} - D_{i,eq}|}{D_{i,eq}} \times 100\%. \tag{10}$$

$N$ represents the number of the samples. $D_{i,ab}$ and $D_{i,eq}$ are the results from first-principles calculations with CI-NEB method and Eq. (9), respectively. The calculated mean deviation $\Delta$ is about 10.3%. Therefore, the method that combining Eq. (9) with the total energy curves of single-Li and two-Li AB-stacked intercalated BLG systems from first-principles calculations, is feasible in estimating the diffusion barriers for multi-Li intercalated AB-stacked BLG systems. Using the diffusion barriers obtained from Eq. (9), we have applied KMC to predict macroscopic diffusion coefficient $D_{Li}$ of Li.

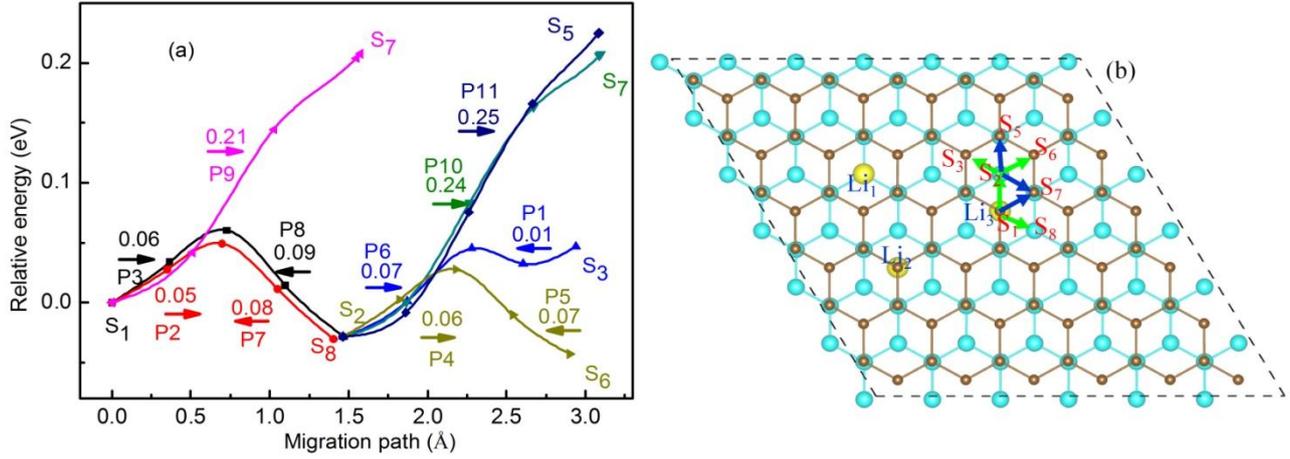

Fig. 8 (a) The total energy profiles and (b) the schematic diagram for three-Li intercalated AB-stacked BLG.

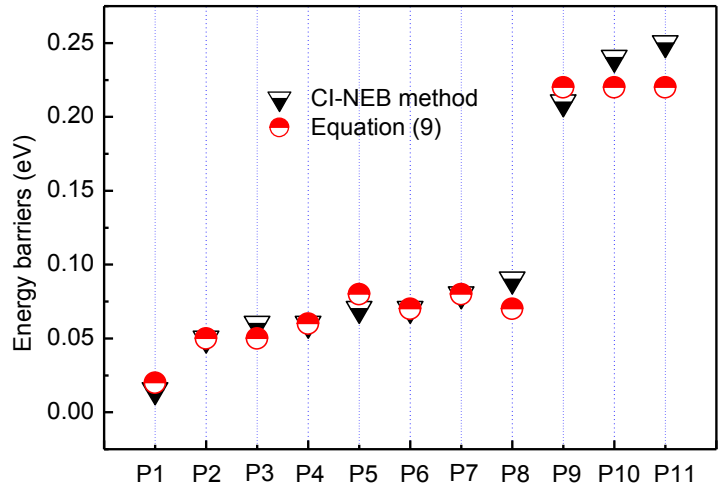

Fig. 9 The energy barriers obtained from CI-NEB method and from Equation (9) for various Li's migration pathways shown in Fig. 8.

Fig. 10 shows the trajectories of Li atoms in the AB-stacked BLG with Li content x=0.06 for T=300K. In the figure, Li diffuses along $TH \rightarrow TH$ pathway instead of $TH \rightarrow TT$ pathway, which is consistent with the result of Fig. 4(a) that Li needs to overcome a much larger energy barrier along $TH \rightarrow TT$ than along $TH \rightarrow TH$ pathway. It confirms the directional preference of Li diffusion more intuitively. The calculated Li diffusion coefficients $D_{Li}$ for the $Li_xC_{12}$ AB-stacked BLG systems with x<0.7 in a temperature range from 263 K to 333 K are presented in Fig. 11(a). The values of $D_{Li}$ are about $0.3 \times 10^{-4} \sim 1.0 \times 10^{-4}$ $cm^2 \cdot s^{-1}$ for the systems with Li content 0.15<x<0.61 near room temperature

(300 K), which explains well the measured value up to $0.7\times10^{-4}$ cm$^2$·s$^{-1}$ of in-plane Li diffusion coefficient within bilayer graphene at room temperature [27]. This good agreement with measured experimental data confirms again that our KMC simulations are reliable. As previously analyzed, the ultrafast diffusion of Li mainly attributes to the stacking structure of AB-stacked BLG which greatly affects its height of potential well within BLG. Furthermore, it is easily found from Fig. 11(a) that the diffusion coefficient $D_{Li}$ decreases as Li content x increases. This accords with the general trend of experimental and theoretical studies on graphitic carbon [12,13]. Furthermore, $D_{Li}$ increases with rising temperature, and the calculated results for Li intercalated BLG with x<0.54 yield a better linear relationship between $\ln(D_{Li})$ and $\frac{1}{k_B T}$, as shown in Fig. 11(b). It indicates that the temperature dependence of $D_{Li}$ could be described by the Arrhenius law [63,64], $D_{Li}(T) = D_0 \exp(-E_A/k_B T)$. Here, $D_0$ is the pre-exponential factor, $E_A$ represents macroscopic activation energy.

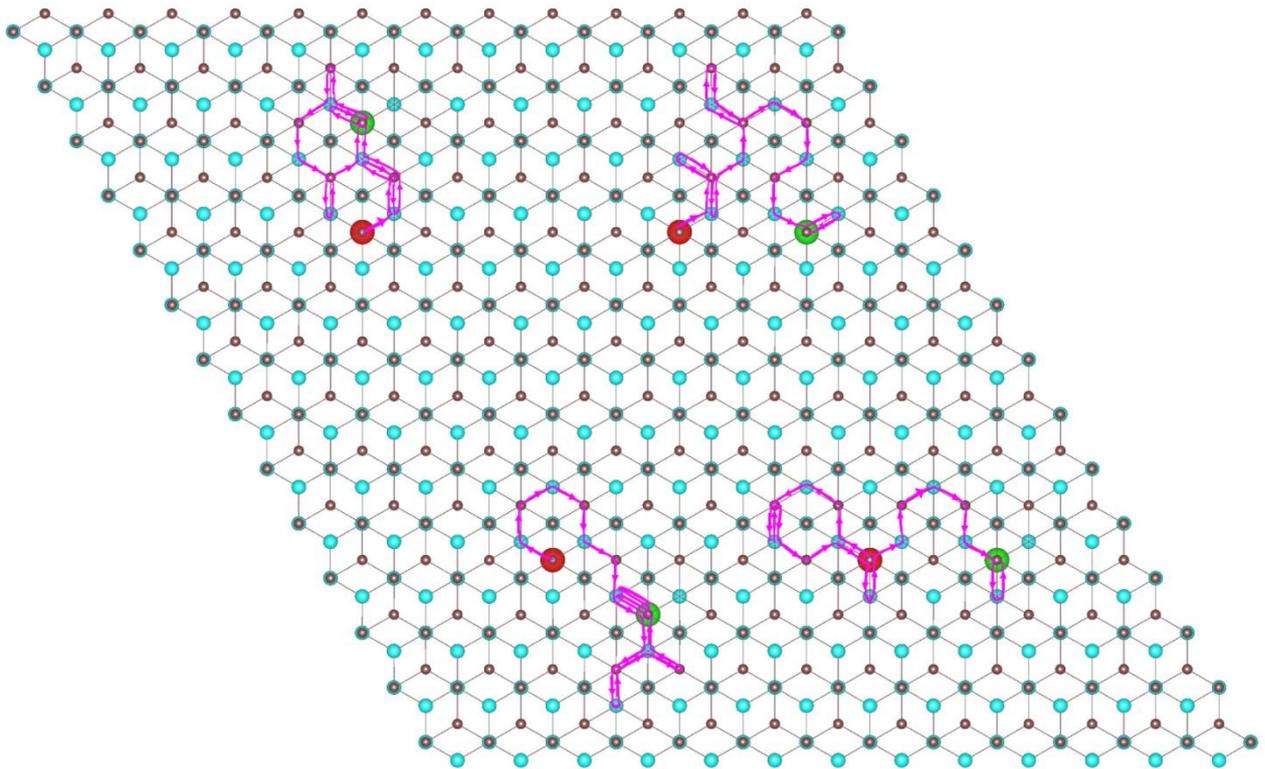

Fig. 10 Trajectories of Li atoms of the AB-stacked BLG with Li content x=0.06 for T=300K, which show 20 steps. Li atoms start at the origin (red dot) and end at the green dot.

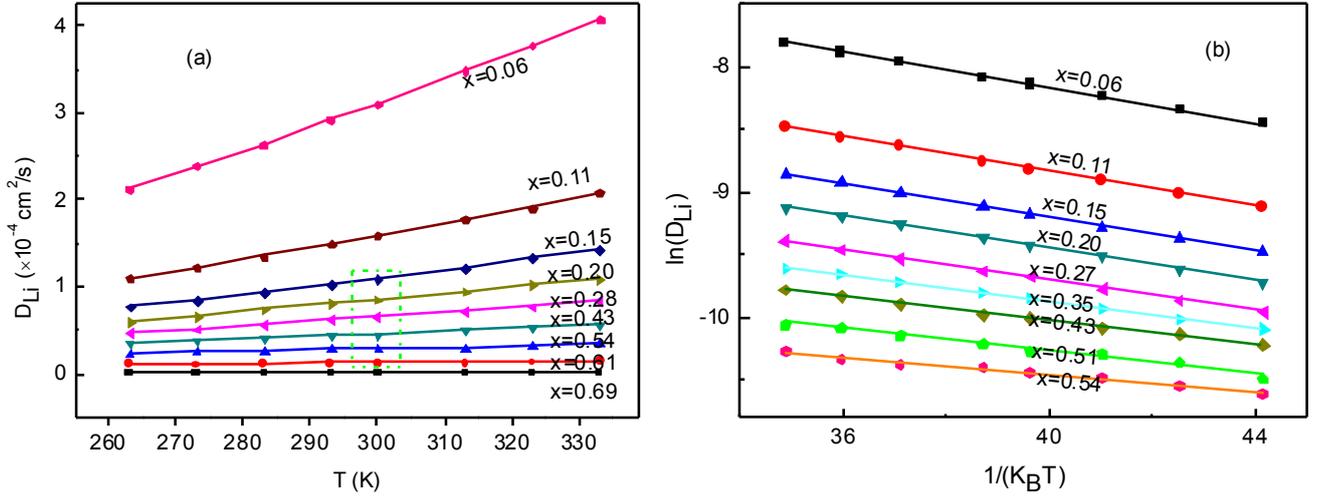

Fig. 11 (a) The diffusion coefficients $D_{Li}$ as a function of temperature $T$, (b) the logarithms of $D_{Li}$ as a function of $1/k_BT$ and their fitting lines, for various Li content x.

## Ⅳ. CONCLUSION

In summary, we have investigated the adsorption and diffusion of Li in the bilayer graphene by combining first-principle calculations with KMC simulations. It is found that Li energetically prefers to intercalate within the bilayer graphene. Li atoms transfer electrons to the BLG so that they become Li ions when they have intercalated within the BLG. Interestingly, Li has an ultrafast diffusion within AB-stacked BLG but not within AA-stacked BLG near room temperature. This may attribute to the stacking structure of AB-stacked BLG which greatly affects its height of potential well within BLG. It implies that changing the stacking structure of BLG is one possible way to greatly improve Li diffusion rate within Li-intercalated BLG. Furthermore, both the stacking structure and the interaction among Li ions cause Li diffusion within AB-stacked BLG to exhibit directional preference. The temperature dependence of Li diffusion could be described by the Arrhenius law. These findings can help the rational design of the anode material in high charge/discharge rate LIBs, also have practical significance for the application of electronic devices based on the Li-intercalated BLG structures.

## ACKNOWLEDGEMENTS

This work was supported by the National Science Foundation of China (61574037, 61404029, 11404058, 11274064), Science and Technology Major Projects of Fujian Province (2013HZ0003), Project of Fujian Development and Reform Commission (2013-577), the Natural Science

Foundations of Fujian Province of China (2016J05151, 2016J01011). G. Xu would like to thank the support from Fujian Provincial College Funds for Distinguished Young Scientists of year 2015.